# *"Phonon Invisibility"* Driven by Robust Magneto-Elastic Coupling in AlFeO$_3$ Thin Film

Shekhar Tyagi, Gaurav Sharma, R. J. Choudhary and Vasant G. Sathe*
*UGC-DAE-Consortium for Scientific Research, Devi Ahilya University Campus, Indore 452001, India*
*Email: vasant@csr.res.in

**Abstract**

The thin films of lead free magneto-electric compound AlFeO$_3$ have been deposited using pulsed laser deposition technique. X-ray diffraction, X-ray absorption spectroscopy and reflectivity measurements established the orthorhombic structure and material density of 4.5g/cc which is comparable with bulk AlFeO$_3$. The Raman mode corresponding to AlFeO$_3$ was found to vanish when magnetic field of 800 Oe was applied at room temperature. Additionally, it was observed that the Raman phonon mode present at the room temperature becomes invisible in the temperature window of 280K-236K and reappears below it. The detailed analysis of magnetization showed a change in magnetic order in this temperature interval. The invisibility of Raman phonon mode corresponding to AlFeO$_3$ have been attributed to the lattice deformation caused by the magnetoelastic effect. The presence of strong spin-lattice coupling is also validated by the renormalization of phonon frequencies below 200 K.

## 1. Introduction

Magnetoelectric (M-E) and multiferroic materials are being explored extensively due to their potential applications as next generation devices by the virtue of mutual control of electric and magnetic properties. The major challenge is to search single phase magnetoelectric materials with suitable properties for practical applications. At present, most of the single phase materials demonstrating magnetoelectric effect contain heavy metals such as lead (Pb), which are hazardous [1]. AlFeO$_3$ (AFO) is a lead-free perovskite material, which shows noticeable magnetoelectric and multiferroic properties [2,3,4,5,6].

The crystal structure of AFO consists of alternate layers of cations (Al / Fe) and O ions stacked along the crystallographic *c* direction with four cationic sites: Fe1, Fe2 and Al2 showing the octahedral environment while the Al1 site exhibits tetrahedral environment. AFO crystallizes in two phases, orthorhombic (Pna2$_1$) phase with collinear ferrimagnetic structure having Neel temperature (T$_N$) between 210–250 K [7] and rhombohedral phase (R-3c) with ferrimagnetic transition T$_N$ at ~225 K [7]. In ferrimagnetic phase, the Fe and Al ions occupy the cationic sites non-preferentially and thus, this structure is known as disordered structure [8]. Recently, Mandar et al. [9] reported that AFO nanoparticles with rhombohedral phase (R-3c space group) show anti-ferromagnetism and weak ferroelectric properties near room temperature. This suggests that the magnetic order is susceptible to changes in size and strain. Thus, AFO can be used as a prospective lead free multiferroic material for the spintronic applications. Yosuke et al. [10] described epitaxial growth of metastable multiferroic AlFeO$_3$ film on SrTiO$_3$ (111) substrate which shows enhanced ferrimagnetic transition temperature, T$_N$ = 317K and pinched-like hysteresis loop along with ferroelectricity at room temperature. Han et al. [11] showed in GaFeO$_3$ which is isostructural to AlFeO$_3$ that nanoparticles with different grain size affects the site disorder and as a consequence, the magnetic properties show a change.

Magnetoelastic coupling is important both scientifically and technologically and is one of the possible mechanisms for inducing magnetoelectric coupling. The magnetoelastic coupling is a phenomenon where the lattice shows strong deformation with the application of magnetic field. If the deformation is sufficiently large it can induce electric polarization. AFO is a magnetoelectric material with different cationic ordering and it shows a first order transformation from orthorhombic phase to rhombohedral phase with large change in volume [7]. The two structurally dissimilar phases show significant variation in magnetic properties [7]. This hints towards the sample to be a promising magnetoelastic material. As mentioned before, thin films of AFO shows magnetic order at room temperature, thus, it is likely to show magnetoelastic coupling at room temperature. Till now there are no reports on the magnetoelastic coupling in the AlFeO$_3$ and similar materials.

Therefore, we have grown polycrystalline thin film of AFO compound on LSAT (001) substrate using pulse laser deposition (PLD) technique. Herein, we report strong lattice deformation on application of magnetic field at room temperature resulting in dramatic modification in the phonon structure of the AFO film. The strong magnetoelastic coupling was also evidenced in temperature dependent Raman spectroscopy study where modification in the magnetic ordering results in diminishing phonon mode.

## 2. Experimental techniques

The AFO nanoparticles were prepared by sol-gel chemical route. The as prepared precipitates were pressed in the form a pellet and sintered at 1300°C to form the target which was used during film deposition using PLD technique. A KrF excimer laser (248 nm) with 5Hz frequency, energy density of about 3 J/cm$^2$ was used to grow a 30-nm polycrystalline thin films on LSAT (001) single crystal substrate. The substrate temperature was kept at 750°C under 0.530 mbar oxygen partial pressure during deposition. In one of the films, a buffer layer of SrRuO$_3$ (SRO) was also deposited on LSAT substrate. For confirming the phase purity, x-ray diffraction was carried out using BRUKER D2 Phaser (Cu K-α radiation) and X-ray reflectivity (XRR), reciprocal space mapping (RSM) measurements were carried out in Bruker D8-Discover high resolution X-ray diffractometer (HRXRD). X-ray absorption spectroscopy (XAS) technique was used for chemical composition analysis in total electron yield mode at O (oxygen) K-edge and Fe (iron) L-edge at the soft x-ray absorption spectroscopy beamline, BL-1[12], Indus-II synchrotron radiation source. The detailed temperature dependent Raman spectroscopic study was carried out using HR-800 Horiba Jobin Yvon micro-Raman spectrometer equipped with He-Ne laser beam (λ=632.8 nm). We have also performed the Raman measurements in the presence of external applied magnetic field of ~ 800 Oe at ambient conditions. The magnetic properties of the film as a function of temperature under 200 Oe magnetic field was examined using 7-Tesla SQUID-vibrating sample magnetometer (SVSM; Quantum Design Inc., USA).

was obtained by depositing AlFeO$_3$ directly on LSAT (001) substrate. The deposition conditions and thickness of the films were kept same for the two films.

Figure 1(a) shows the X-ray diffraction pattern of the (A1) and (A2) along with the substrate LSAT (001) collected in θ-2θ mode. Both the films show a single broad peak corresponding to AFO apart from peaks related to substrate and SRO. This suggested that the films are highly oriented. We attempted reciprocal space map on these films but failed to observe a reflection corresponding to the film. This indicates that the films are not epitaxial in nature. The film peak position nearly matched with the (022) AFO peak in orthorhombic structure [JCPDS card No. 84-2153]. The observed broadness of the reflection corresponding to the AFO layer is attributed to the reduced crystallite size and to the thinness of the films. The film thickness and material density was calculated using XRR measurements carried out on the AFO thin films (not shown here). This resulted in 30 nm thickness and 4.5g/cc material density of the films which matches with the bulk AFO. In addition, the epitaxial nature of the SRO buffer layer is confirmed by RSM study [not shown here].

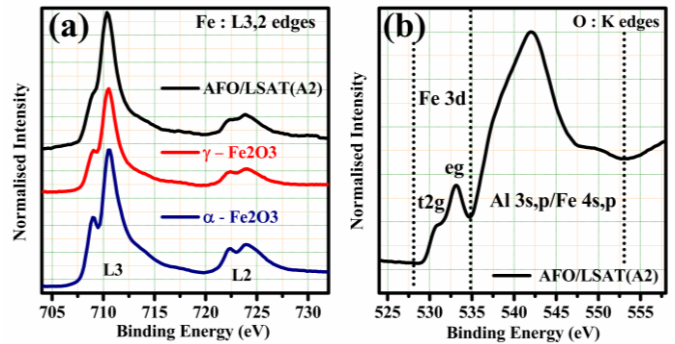

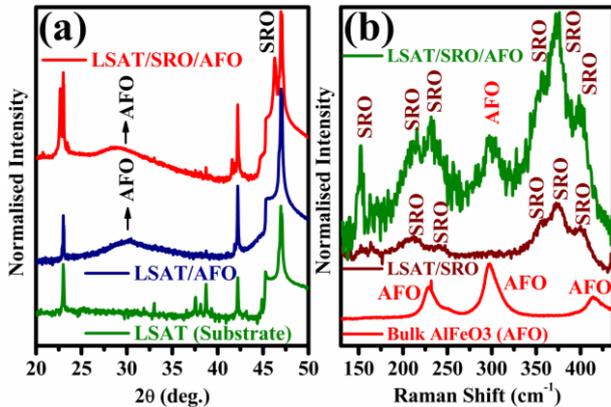

**Figure 1.** (a) The recorded X-ray diffraction patterns for AlFeO$_3$/SrRuO$_3$/LSAT(001), AlFeO$_3$/LSAT(001) and pure LSAT(001) substrate. (b). The Raman spectra collected on AlFeO$_3$/SrRuO$_3$/LSAT(001), SrRuO$_3$/LSAT(001) and bulk AlFeO$_3$.

## 3. Results and discussions

In this study, two films were investigated in detail. The first film (A1) is obtained by depositing AlFeO$_3$ on LSAT (001) substrate with a SrRuO3 buffer layer AlFeO$_3$/SrRuO$_3$/LSAT(001) while the second film (A2)

**Figure 2.** (a) XAS spectrum of AlFeO$_3$/LSAT (A2) film recorded at 300 K along with the XAS of reference samples of iron oxides, α-Fe$_2$O$_3$ and γ-Fe$_2$O$_3$ collected at Fe L edges (b) Oxygen K-edge, XAS spectra of AlFeO$_3$/LSAT (A2) film at 300 K.

Figure 1(b) shows the recorded Raman spectra corresponding to the bulk AFO compound, SRO/LSAT (001) film (deposited for comparison) and the A1 film. The Raman modes of the AFO deposited directly on the LSAT substrate (A2) were found to be buried in the substrate signal, however, the Raman mode is observable on film (A1) where SRO is deposited as a buffer layer (see figure 1(b)). The bulk AFO showed three prominent Raman modes and the overall Raman signatures matched with the previous reports [7]. Out of the three Raman modes, the mode occurring at ~ 315 cm$^{-1}$ is clearly visible in A1 film, the wavenumber of the other two modes clashes with the wavenumber of the SRO modes.

For confirming the chemical environment, we have carried out x-ray absorption spectroscopy study on the A2 film. Figure 2(a) shows the XAS spectrum at Fe: $L_{3,2}$ edges of the A2 film along with the XAS spectrum recorded on iron oxides, $\alpha$-$Fe_2O_3$ and $\gamma$-$Fe_2O_3$. It is well known that the $\alpha$-$Fe_2O_3$ possesses only octahedral $Fe^{+3}$ sites while the $\gamma$-$Fe_2O_3$ possesses a mixture of tetrahedral and octahedral $Fe^{+3}$ sites [13].

When compared, the spectral features at $L_2$ and $L_3$ edges of A2 film are found to be closer to the spectral features of $\gamma$-$Fe_2O_3$ rather than that of $\alpha$-$Fe_2O_3$. This indicates that the Fe cations occupy both the tetrahedral and octahedral sites in A2 film. The presence of mixture of cationic sites suggests the disordered structure in A2 film at 300 K. Figure 2(b) shows the XAS spectrum collected at O: K edge of A2 film. The spectral features at low binding energy corresponds to the $t_{2g}$ and $e_g$ states induced by Fe 3d while high energy features are induced by Al 3s;p and Fe 4s;p hybridized states. These spectral features exactly match with the spectral features reported on single crystal $GaFeO_3$ which show $AlFeO_3$-type orthorhombic crystal structure [14].

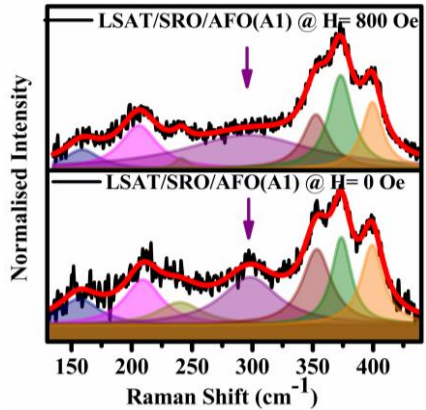

**Figure 3.** The Raman spectra recorded at 300 K in the presence (H= 800Oe) and absence (H= 0Oe) of magnetic field on $AlFeO_3$/$SrRuO_3$/LSAT (001) thin film (A1).

In order to investigate magnetoelastic coupling at room temperature, we have recorded the Raman spectra in the presence of applied external magnetic field (H= 800 Oe) on A1 film at room temperature that is shown in figure 3 and in order to facilitate comparison, Raman spectra recorded without magnetic field is also presented along with the fitted profiles using Lorentzian function. Interestingly, on the application of field, the Raman mode of the AFO diminishes while all other Raman modes remained unaffected. This is a direct evidence of strong lattice deformation induced by the external magnetic field. As mentioned before, at room temperature this compound shows a ferrimagnetic behavior, on the application of field, the ferrimagnetic order is getting affected which in-turn produces strong lattice deformation. This modulation in lattice due to changes in magnetic order is reflected in the form of annihilation of the phonon mode. It would have been very interesting to observed changes in crystal structure under magnetic field, however, due to the small film thickness the only x-ray diffraction peak observed is extremely broad and hence it is not likely to provide any meaningful information.

The Raman spectra was also collected as a function of temperature in the temperature range of 360K to 83K for the A1 film. The selected Raman spectra are illustrated in figure 4(a). We focused our attention on the 315 $cm^{-1}$ mode corresponding to the AFO. It is seen that this mode is present in high temperature range marked by an arrow and then suddenly disappears below 280 K. It remains absent upto 230 K (see spectra marked by a rectangle in the figure). Surprisingly, the mode reappears at and below 230 K (marked by a circle). At low temperatures, this mode showed asymmetric shape profile. The Raman modes of AFO and SRO are fitted using Lorentzian function and the resulted mode positions as a function of temperature are plotted in figure 4(b). Vertical dotted lines in figure 4 (b) indicates the temperature window of 236 to 280K, where the Raman mode of AFO is absent. It is worth noting here that the Raman modes corresponding to the SRO are present in this temperature window and showed normal behavior. This thus, eliminates the possibility of experimental artifact.

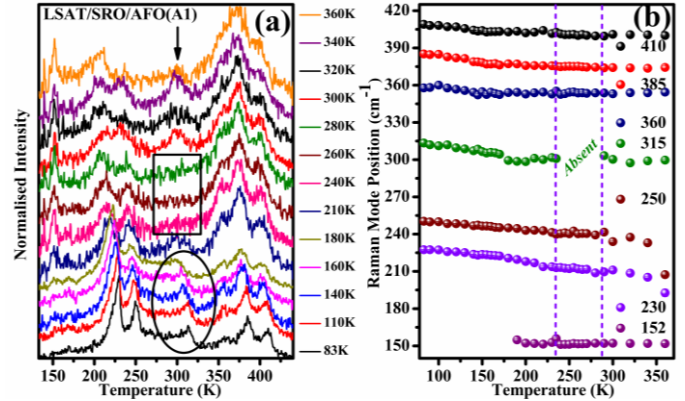

**Figure 4.** (a) Temperature dependent Raman spectra of $AlFeO_3$/$SrRuO_3$/LSAT(001) film. Rectangle and circle indicates absence and re-entrance of the Raman mode corresponding to AFO, respectively. (b) The Raman mode wavenumbers as a function of temperature. Vertical lines indicate transition region.

The measurement was repeated again and similar behavior was observed. The invisibility of phonon is similar to that observed in the Raman spectra recorded under field at room temperature. This suggests magnetic origin to this phenomenon.

In order to find out magnetic properties of the film, temperature dependent magnetization (M-T) study was carried out in 5-350K temperature range under field cooled (FC) and zero field cooled (ZFC) protocol on A2 film that are shown in figure 5 (a). When cooled, it showed a broad maxima before showing a downward trend in ZFC while the FC showed systematic increase. In bulk compound this system shows ferrimagnetism with $T_N \sim 250$ K [7] while in epitaxial thin films a ferrimagnetic transition at ~317 K is reported with similar magnetization behaviour [10].

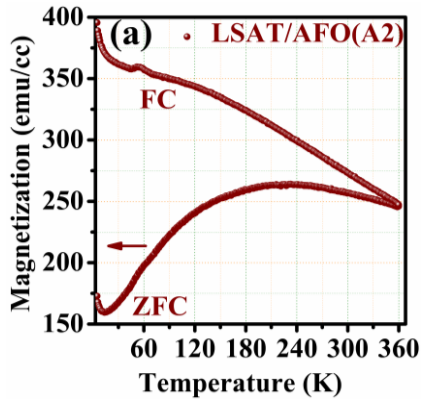

**Figure 5.** (a). Magnetization of AlFeO$_3$/LSAT(001) film as a function of temperature measured under 200 Oe in field cooled (FC) and zero field cooled (ZFC) protocol.

From the ZFC, the inverse of DC susceptibility was extracted and is plotted as a function of temperature in Figure 6 (a). Similarly, the magnetic anisotropy ($\chi_{ZFC}$-$\chi_{FC}/\chi_{FC}$) as a function of temperature is also deduced from the magnetization measurements and is presented in Figure 6 (a). Interestingly, the extrapolation of the slope of the $1/\chi_{ZFC}$ vs T clearly shows a deviation from the linear behaviour at 280 K and $\chi_{ZFC}$-$\chi_{FC}/\chi_{FC}$ curve shows a discontinuous change in the slope at 236 K. This suggests that the magnetic order and magnetic anisotropy renovates in this temperature range. This is a direct evidence for the magnetic origin of the phonon invisibility. In figure 6 (b), the phonon frequency and the full width at half maxima (FWHM) for (315 cm$^{-1}$) AFO Raman mode is plotted as a function of temperature. It shows a strong modulation in the temperature window where magnetic ordering also showed perturbation. This establishes presence of strong magneto-elastic coupling in AFO thin film.

As mentioned before, asymmetric line shape in the AFO Raman mode (315 cm$^{-1}$) profile is observed below 200 K along with phonon frequency renormalization (see figure 6 (b)). The renormalization of phonon mode provides an evidence of the presence of strong spin-lattice coupling. The asymmetric line shape can arise due to number of reasons. This compound is highly insulating and hence the possibility of electronic scattering is scarce. The asymmetry can thus arise due to lattice deformation which is likely to occur in compounds showing strong lattice modulation due to magnetic order.

The annihilation of phonon mode due to application of magnetic field suggests that the magnetic order observed in the temperature window 280 K - 236 K is reproduced by application of magnetic field at room temperature. The strong magnetoelastic coupling observed in the present study at room temperatures and moderate field (800 Oe) opens up possibility of usage of these thin films in device applications such as: sensors, magnetic memory devices, spintronic devices, etc. The orthorhombic space group Pna21 supports pyroelectricity and hence there exists a finite probability of contribution to the electric polarization induced by the spin-order in this multiferroic thin film due to the combined effects of magnetoelastic coupling and piezoelectric effect which need to be explored further.

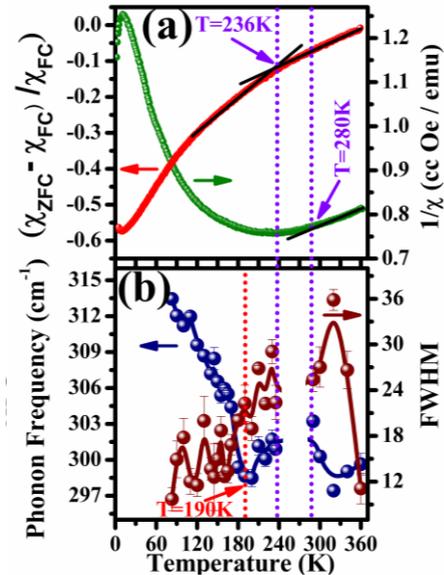

**Figure. 6.** (a). Inverse susceptibility ($1/\chi_{ZFC}$) and the magnetic anisotropy ($\chi_{ZFC}$-$\chi_{FC}/\chi_{FC}$) as a function of temperature. (b). Phonon frequency and full width at half maxima of the AFO mode as a function of temperature. Vertical lines represents transition region.

## 4. Conclusions

In conclusion, AlFeO$_3$ showed dramatic changes in phonon structure under application of 800 Oe magnetic field at room temperature, the phonon mode corresponding to AlFeO$_3$ diminishes completely on the application of field in Raman spectroscopy study. The phenomenon is explained by considering large magnetoelastic coupling induced lattice deformations. Temperature dependent Raman spectroscopy showed invisibility of the Raman mode corresponding to the AFO phase in the temperature interval of 280 K–236 K where the inverse of susceptibility and magnetic anisotropy function showed modulation. This reconfirmed the existence of strong magnetoelastic coupling in this material. Below 200 K, the re-normalization of phonon frequency along with asymmetric line shape is observed, which further provided strong evidence of the presence of magneto-elastic coupling. Such a strong magnetoelastic behavior at room temperature can be useful in many device applications.